\documentclass[twocolumn]{revtex4-1}
\usepackage{graphicx}
\usepackage{amsmath}

\begin{document}

\title{Two channel model for optical conductivity of high mobility organic crystals}

\author{A. de Candia$^{1,2}$, G. De Filippis$^{1,2}$, L.M. Cangemi$^2$, A.S. Mishchenko$^{3,4}$, N. Nagaosa$^{3,4}$ and V. Cataudella$^{1,2}$}

\affiliation{
$^1$Dip. di Fisica - Universit\`{a} di Napoli "Federico II" - Comp. Univ. di Monte S. Angelo I-80126 Napoli, Italy \\
$^2$SPIN-CNR Comp. Univ. di Monte S. Angelo I-80126 Napoli, Italy \\
$^3$RIKEN Center for Emergent Matter Science (CEMS), Wako, Saitama 351-0198, Japan\\
$^4$Department of Applied Physics, The University of Tokyo 7-3-1 Hongo, Bunkyo-ku, Tokyo 113-8656, Japan
}


\begin{abstract}
We show that the temperature dependence of conductivity of  high mobility organic crystals 
Pentacene and Rubrene can be quantitatively described in the framework of the model 
where carriers are scattered by quenched local impurities and interact with phonons by 
Su-Schrieffer-Hegger (SSH) coupling.
Within this model, we present approximation free results for mobility and optical conductivity obtained
by world line Monte Carlo, which we generalize to the case of coupling both to phonons and
impurities. 
We find fingerprints of carrier dynamics in these compounds which differ from conventional metals and
show that the dynamics of carriers can be described as a superposition of a Drude term representing 
diffusive mobile particles and a Lorentz term associated with dynamics of localized charges.
\end{abstract}

\maketitle

\section{Introduction}
Numerous technological applications \cite{hasegawa,morpurgo_1,bredas,mei,MENG201733}
of  high mobility organic crystals Pentacene and Rubrene attracted considerable interest to its transport properties.
However, in spite of intense efforts, there are only few theoretical approaches capable of providing 
reasonable descriptions of the physical properties of these compounds \cite{troisi_1,troisi_2,fratini_1,fratini_1b,cataudella_1,ciuchi_1,fratini_2,fratini_3,fratini_4,noi}. 
Moreover, a complete description of experiments is achieved only in the high-temperature range 
$150\,\mbox{K}<T<350\,\mbox{K}$.
In this paper we present a novel model which is able to provide a unified description 
of the transport properties of these compounds in the whole temperature range.
The model is based on the traditionally used Su-Schrieffer-Hegger model, which is generalized 
in our study by adding a coupling of carriers to quenched local impurities.
We solve this model by a novel approximation free modification of the world line Monte Carlo (WLMC)
approach, demonstrate agreement with experimental mobility at all temperatures, and give 
physical interpretation of the optical conductivity (OC) and diffusivity of these compounds.

To review the current theoretical understanding, we start with the pioneering approach by 
Troisi-Orlandi \cite{troisi_1,troisi_2}, using the one dimensional (1D) Su-Schrieffer-Hegger (SSH) 
model, where optical phonons of very low energies ($\hbar\omega_0\sim\,\mbox{few meV}$) 
are coupled to the electrons characterized by a hopping of the order of $t_h\simeq 100\,\mbox{meV}$,
so that the relation $t_h>k_BT\gg\hbar\omega_0$ holds between the model parameters.

Indeed, the mobility in these materials is characterized by a pronounced anisotropy (Ref. \cite{morpurgo_1} and references therein),
so that a one dimensional model, where carrier hopping in the other two directions is completely neglected,
is a good approximation. Note that this is the main approximatoin of our approach, as we extract the mobility using a method which is approximation free within the model Hamiltonian chosen.

Actually a more detailed description of these semiconductors would require a
contribution \`a la Holstein, that is a local electron-phonon coupling, describing the
interaction between the charge carriers and the intramolecular deformations.
On the other hand, in literature, the characteristic frequency of these
vibrations has been estimated to be of the order of 1-2 $t_h/\hbar$ \cite{bredas,bredas_2},
much greater than the frequency associated to the intermolecular displacements, modelled by means of the SSH Hamiltonian,
that is about 0.05 $t_h/\hbar$.
The low energy physics, at $k_BT$ much less than $t_h$, is determined mainly by the coupling between the electrons and the lattice excitations
within the SSH model (nonlocal electron-phonon interaction). Note that $t_h/k_B$ is about 1200 kelvin.

In literature, the 1D model has been studied by using different approximate approaches, reproducing the experimental power law $\mu\sim T^{-\gamma}$
temperature dependence, where the exponent $\gamma$ depends on materials and disorder \cite{morpurgo_1}. 
These approaches are based on:
(i) mixed quantum-classical simulations based on the Ehrenfest coupled equations \cite{troisi_1,ciuchi_1};
(ii) neglecting vertex corrections in the OC calculation \cite{fratini_1,fratini_1b};
(iii) introducing an ad hoc energy broadening of the system energy levels \cite{cataudella_1};
(iv) using the relaxation time approximation \cite{ciuchi_1,fratini_2,fratini_3};
and (v) methods based on realistic chemical models \cite{landi_1,landi_2}.

The scenario emerging in these approaches, i.e. the ``transient localization scenario'', has been largely adopted to describe these materials \cite{fratini_4}.

It assumes that, on a time scale less than $\tau \approx 1/\omega_0$, the electrons tend to localize due
to Anderson theorem \cite{anderson_0} (the lattice fluctuations are considered "frozen" on this time scale),
while for times larger than $\tau > 1/\omega_0$ the electrons acquire a diffusion.

Recent approximations free study \cite{noi} of the 1D SSH Troisi-Orlandi model confirmed power law of mobility 
temperature dependence, and established ultimate understanding of the high-temperature behavior of mobility.

However, the approaches based solely on the SSH model \cite{troisi_1,troisi_2,fratini_1,noi,ciuchi_1} 
are unable to describe low temperature ($T<180 K$) transport of Pentacene and Rubrene because 
this model does not account for static extrinsic disorder of organic crystals 
\cite{morpurgo_1,C7MH00091J,Mishchenko_disorder0,Mishchenko_disorder,sergio_disordine}.
To resolve this drawback of disorder-free scenario we add the effect of quenched static local impurities
into 1D SSH Hamiltonian, to end up with the model giving comprehensive description of the
$\omega$-dependent OC $\sigma(\omega,T)$ and mobility of the crystals under study.
Rewriting long-established WLMC \cite{kornilovich} into momentum space, and generalizing it to 
the interaction of polaron with impurities, we use analytic continuation to real frequencies based 
on Maximum Entropy \cite{JM} and Stochastic Optimization \cite{andrey} methods
to obtain OC from the current-current correlation function.  
Our analysis 
reveals two contributions to mobility and OC:
one is associate with the diffusive motion of a mobile quasiparticles, and the second arises from the 
excitations of  localized carriers.

\section{Model and methods}
The model Hamiltonian is given by $H=H_1 + H_2$, in momentum space
\begin{equation}
H_1=\hbar\omega_0\sum\limits_k a_k^\dag a_k + \frac{1}{\sqrt{N}}\sum_{q,k}\delta_{k} c_{q+k}^\dag c_q\; 
\label{h_q}
\end{equation}
and
\begin{equation}
H_2 = \sum_{q}\varepsilon(q)c_q^\dag c_q +
\frac{1}{\sqrt{N}}\sum_{q,k} M(k,q)
\left(a_{-k}^\dag+a_{k}\right)c_{q+k}^\dag c_q \;,
\label{he1_q}
\end{equation}
where operators $c_{k}^{\dagger }$ ($a_{k}^{\dagger }$) represent the electron  (phonon) creation operators,
Einstein optical phonons have frequency $\omega_0$, and ${\delta_k}$
are random variables representing the static on site disorder, distributed according to a Gaussian with $\langle \delta_k\rangle=0$,
and $\langle\delta_k\delta_{k^\prime}\rangle=\Delta^2$ if $k^\prime=-k$, zero otherwise.
The electron band is $\varepsilon(q)=-2t_h\cos(q)$, with $t_h$ the hopping parameter, while $M(k,q) =2it_h\lambda\left[\sin q-\sin(q+k)\right]$
is the EPI vertex, with $\lambda$ the dimensionless coupling constant.

We study the model with a WLMC approach, where the trace over the state vectors (we work in the thermodynamic limit $N\to\infty$) becomes a sum over 
all the paths in imaginary times (world lines). 
A world line is defined by a function $q(\tau)$ of the electron momentum,
piecewise constant and periodic in the interval $[0,\beta]$. The function has a finite number of discontinuities (hoppings) due to the interaction with phonons and with disorder.
Each hopping is paired with a corresponding one having opposite momentum. 

It is convenient to work in the Suzuki-Trotter approximation, in which the operator $e^{-\beta H}$ is written as the product of a finite number $L$ of ``time slices'',
$\left[e^{-(\beta/L)H}\right]^L$, and take the limit $L\to\infty$ at the end. The partition function, after insertion of the appropriate number of identities, can be written as
\begin{equation*}
Z=\sum_{|\phi_1\rangle}\cdots\sum_{|\phi_L\rangle}\langle\phi_1|e^{-(\beta/L)H}|\phi_2\rangle\langle\phi_2|e^{-(\beta/L)H}|\phi_3\rangle\cdots
\end{equation*}
The vectors $|\phi_k\rangle$ represent the state of the system (electron + phonons) at imaginary time $\tau_k=(\beta/L)k$, with $k=1\ldots,L$.
The electron kinetic term of the Hamiltonian gives simply
$\exp[2t_h\int\limits_0^\beta\!\cos q(\tau)\,d\tau]$.
Of course, before taking the limit $L\to\infty$, the integral over $\tau$ has to be considered as a sum over the discrete times $\tau_i$.
If the world line has $n$ hoppings due to phonons, let us call
$\tau_{2i-1}<\tau_{2i}$ the imaginary times at which the hoppings occur, with $i=1,\ldots,n/2$ and $n$ even,
$q_{2i-1}$ and $q_{2i}$ the momentum of the electron just before the hopping, and $k_{2i-1}$ and $k_{2i}$ the momentum of the phonon emitted or absorbed,
with $k_{2i}=-k_{2i-1}$. While diagonal terms in the Hamiltonian contribute to the product over all the time slices where the electron does not hop,
non-diagonal terms like electron-phonon interaction $H_{\text{epi}}$ contribute only when the electron hops. 
Therefore, for $L\to\infty$, the term $\exp[-(\beta/L)H_{\text{epi}}]$ can be replaced by $1-(\beta/L)H_{\text{epi}}$.
The matrix elements of this operator, and of the diagonal operators $\exp[-(\beta/L)\hbar\omega_0a_k^\dag a_k]$ relative to
the phonon being emitted or absorbed over the whole interval $[0,\beta]$, are given by
\begin{align*}
&\frac{\beta^2}{NL^2}M(k_{2i-1},q_{2i-1}) M^*(k_{2i},q_{2i})\times
\\
&\left[(n_k+1)e^{-\hbar\omega_0(\beta n_k+|\tau_{2i}-\tau_{2i-1}|)}
+n_k e^{-\hbar\omega_0(\beta n_k-|\tau_{2i}-\tau_{2i-1}|)}\right],
\end{align*}
where $n_k$ is the number of phonons before time $\tau_{2i-1}$, and the two terms correnspond respectively to emission or absorption.
Summing over the number of phonons, apart from a factor $(1-e^{-\beta\hbar\omega_0})^{-1}$ that corresponds to the partition function of free phonons and can be dropped, we have
\begin{equation*}
\frac{\beta^2}{NL^2}M(k_{2i-1},q_{2i-1}) M^*(k_{2i},q_{2i})\frac{\cosh\left[\hbar\omega_0\left(\frac{\beta}{2}-|\tau_{2i}-\tau_{2i-1}|\right)\right]}{\sinh(\beta\hbar\omega_0/2)}
\end{equation*}
Note that the weight of world lines in which more than one phonon with the same momentum is emitted or absorbed vanishes as $N^{-2}$ or faster, and as we will see later is negligible
in the thermodynamic limit.
Calling $k_{2j-1}$ and $k_{2j}$, with $j=1,\ldots,m/2$ and $m$ even, the hoppings of the electron due to the disorder term of the hamiltonian, the matrix elements
over a pair of hoppings 
can be computed similarly, and give $\frac{\beta^2}{NL^2}\delta_{k_{2j-1}}\delta_{k_{2j}}$. Putting all together, we obtain that the weight of the world line is given by
\begin{equation*}
\tilde W[q(\tau)]=\frac{n!!m!!}{N^{(n+m)/2}L^{n+m}}W[q(\tau)],
\end{equation*}
where
\begin{align*}
W[q(\tau)]=\exp\left[2t_h\int\limits_0^\beta\!\cos q(\tau)\,d\tau\right]\frac{\beta^m}{m!!} \frac{\beta^n}{n!!}
\times \\
\prod_{j=1}^{m/2}\delta_{k_{2j-1}}\delta_{k_{2j}}\prod_{i=1}^{n/2}
M(k_{2i-1},q_{2i-1}) M^*(k_{2i},q_{2i}) \times
\\
\frac{\cosh\left[\hbar\omega_0\left(\frac{\beta}{2}-|\tau_{2i}-\tau_{2i-1}|\right)\right]}{\sinh(\beta\hbar\omega_0/2)}.
\end{align*} 
Monte Carlo updates change the values of the hopping momenta, times, and the number of hoppings.
When doing the Monte Carlo simulation, we have to impose the detailed balance condition
\begin{equation*}
A(x\to y)R(x\to y)\tilde W[x]=A(y\to x)R(y\to x)\tilde W[y],
\end{equation*}
where $x$ and $y$ are two different world lines, $A(x\to y)$ is the probability to propose the Monte Carlo move $x\to y$, and $R(x\to y)$ is the probability to accept it.
If for example $x$ is a world line with $n$ hoppings due to phonons, and $y$ a world line with $n+2$ hoppings, then
$A(x\to y)=N^{-1}L^{-2}$, because we have to choose a value of the momentum transferred and two values of the imaginary times of emission and absorption,
and $A(y\to x)=\frac{2}{n+2}$, because we have to choose which pair of hoppings to remove. Therefore, it is not difficult to see that we can choose $R(x\to y)$ such that
\begin{equation*}
R(x\to y)W[x]=R(y\to x) W[y].
\end{equation*}
In this way, all the factors $N$ and $L$ have disappeared, and we can take the limit $L\to\infty$ and $N\to\infty$,
extracting momenta from the whole interval $[-\pi,\pi]$, and times from $[0,\beta]$.
Note that, if a world line contains more than a pair of hoppings with the same value of the absolute momentum, then a factor $N^{-1}$ will still appear in $W[y]$, and the weight will vanish
in the thermodynamic limit. This corresponds to the fact that it is not possible to extract two times the same real number from a continuous interval.
Therefore, the simulation of the $N\to\infty$ system is actually {\em simpler} than that of the finite system, because we have not to deal with different possible pairings of the hoppings.
Note also that, despite we are doing a quenched average over the disorder,
the factor $\delta_{k_{2j-1}}\delta_{k_{2j}}$ can be replaced by its average $\Delta^2$.
Indeed, in the thermodynamic limit, one can average the weight of a world line over an arbitrarily small interval around
$k_{2j-1}$ and $k_{2j}$, that contains an infinite number of possible momenta. This means in practice that quenched and annealed averages coincide.

Within the approach described in Ref.\ \cite{MobHol},
we determine mobility 
$\mu=\lim_{\omega\mapsto 0} \sigma(\omega)/e$ from the OC $\sigma(\omega)$
which is obtained from current-current correlation function
(see Supplemental Material \cite{sm} that includes Refs. \nocite{antonio,sha}). 

\section{The mobility}
\begin{figure}
\begin{center}
\includegraphics[height=5cm]{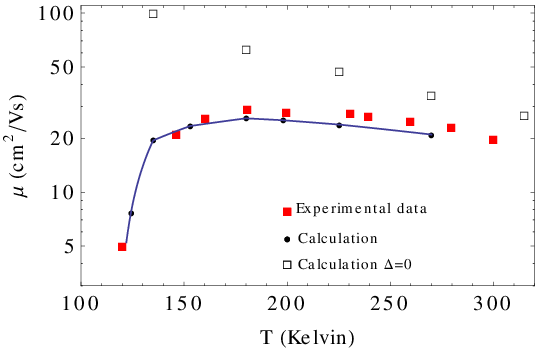}
\caption{Mobility versus temperature: comparison between experimental data (red squares) and calculation (blue points).
The parameters used are:
$t_h=155.0\,\mbox{meV}$, $\hbar\omega_0=7.75\,\mbox{meV}$, $\lambda=0.092$ and $\Delta=38.7\,\mbox{meV}$.
For comparison we plot also the results of the calculation for $\Delta=0$, that is in absence of disorder (empty squares).
Experimental data are taken from Ref.~\cite{morpurgo_1}, Fig.\ 11, data relative to b-axis.}
\label{comparison}
\end{center}
\end{figure}

Our results reproduce the experimental data for rubrene \cite{morpurgo_1} in 
the entire range of temperatures $100\,\mbox{K}<T<300\,\mbox{K}$ (Fig.\ \ref{comparison}), supporting the 
assumption that  the interaction with low energy phonons and static on site disorder are the key ingredients 
needed to describe the experiments, and showing that a finite coupling with quantum phonons is able to 
overcome the Anderson localization and provide  a non-vanishing mobility \cite{bonka}.

At low $T$ the disorder plays the major role, driving the system from metallic to insulating behavior
due to localization on impurities. 
The effect of disorder is very important even in metallic regime at $T>180K$,
where a power law $\mu\propto T^{-1}$ is observed.
Note that in the present context, where the carrier density is very small, with ``metallic regime'' we intend to refer to the
region where the mobility exhibits the typical behavior of the metals, i.e.
power-law temperature dependence: the mobility decreases by increasing the
temperature with an exponent depending on the material and the disorder.
On the other hand with insulating regime we refer to the temperature
window where mobility increases by increasing temperature.
Depending on the disorder strength, the apparent exponent of the power law ($\mu\propto T^{-\gamma}$) can 
vary from $\gamma=1$ (see Fig.~\ref{comparison}) to $\gamma=2$ (negligible disorder), which is in agreement with experimental situations in 
different samples \cite{morpurgo_1}.  

\begin{figure}
\begin{center}
\includegraphics[height=5cm]{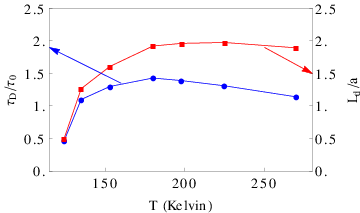}
\caption{Drude scattering time, $\tau_D/\tau_0$, and diffusion length, $L_d/a$, as a function of temperature $T$.}
\label{fig2}
\end{center}
\end{figure}

To identify the transport regime in high mobility organic crystals Pentacene and Rubrene, we estimated 
the diffusion time $\tau_D$ and the diffusion lenght $L_d$, using the exact relation 
$\mu=- e a^2\hbar^{-2}\tau_D\left< H_{2}\right>\label{mu}$ provided by Mori formalism \cite{mori},
and the Einstein equation $\mu=e D/ (k_B T)$, with $D=L_d^2/(2 \tau_D)$.
Here $e$ is the electron charge, $a$ the lattice constant, $k_B$ the Boltzmann constant, 
and $\left<H_{2}\right>$ is straightforwardly obtained from the WLMC calculations.
Figure~\ref{fig2} shows that, in the entire temperature range, the diffusion time $\tau_D$ is of the order of $\tau_0=\hbar/t_h$, which is many times smaller than that in typical 
metals, and its temperature dependence is very similar to that of mobility, pointing to
diffusive behavior.
The diffusion length is of the order of the lattice parameter $a$, and even smaller than $a$ at low temperatures 
(Fig.~\ref{fig2}), violating the Ioffe-Regel condition \cite{IR} and showing that the mobility is dominated 
by a strong incoherent regime. 
Such small $L_d$ corresponds to diffusivity $D=L_d^2/(2 \tau_D)$ which is two order of magnitude smaller
than in typical metals.

\begin{figure}
\begin{center}
\includegraphics[height=9cm]{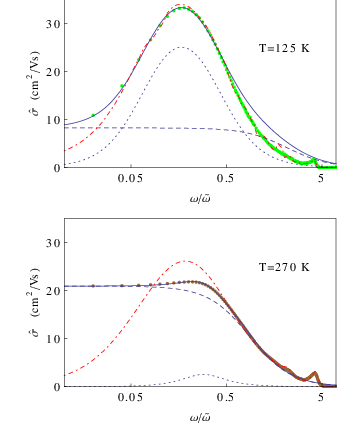}
\caption{Frequency dependent mobility, $\hat{\sigma}=\sigma/e$, for different temperatures (points). The 
solid lines are the best fit in the 
frequency window $0<\omega<{\tilde\omega}/2$, where ${\tilde\omega}=t_h/\hbar$ (see Eqs.\ (\ref{eq_Drude}) and (\ref{eq_Lorentz})), the dashed curves are the Drude contributions and the dotted line 
are the Lorentz contributions. The red dot-dashed line represents the fully adiabatic conductivity.}
\label{conductivity_1}
\end{center}
\end{figure}
  
\section{Optical conductivity: the two channel model}
The profound difference of the low- and high-temperature nature of mobility can be 
highlighted by the analysis of the dynamic OC in the frequency range $0<\omega<{\tilde\omega}/2$, where ${\tilde\omega}=t_h/\hbar=1/\tau_0$ and OC can be 
represented as a linear combination of two contributions (Fig.~\ref{conductivity_1}).
The first one is a Drude term describing the OC of mobile particles at low frequency
\begin{equation}
\sigma_D(\omega)=\frac{\sigma_D}{1+(\tau_D\omega)^2} \; ,
\label{eq_Drude}
\end{equation}
and the second one is the Lorentz term describing the dynamical conductivity associated with localized charges
\begin{equation}
\sigma_L(\omega)=\frac{\sigma_L(\omega\tau_L)^2}{(1-\frac{\omega^2}{\omega_L^2})^2+(\omega\tau_L)^2} \; .
\label{eq_Lorentz}
\end{equation}
Furthermore, for comparison, we show in Fig.~\ref{conductivity_1} also the fully adiabatic conductivity, whose limit for $\omega\mapsto 0$ is zero,
as expected by the Anderson theorem. It is worth noticing that the peak in the exact and fully adiabatic approaches are very close.
Actually for low temperature the agreement is perfect, whereas at room temperature the exact calculation peak is located at a slightly higher frequency,
and exhibits a smaller intensity. This signals the presence of quantum effects even at this high temperature.

\begin{figure}
\begin{center}
\includegraphics[height=5cm]{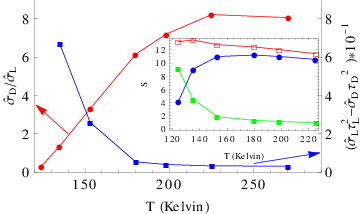}
\caption{The temperature dependence of the  ratio between Lorentz and Drude coefficients $\sigma_D/\sigma_L$, and the temperature dependence of the coefficient of 
the small $\omega^2$ expansion of the sum of Lorentz and Drude optical conductivities. Times are measured in terms of $\tau_0=\hbar/t_h$ and $\hat{\sigma}$ in 
units of $ea^2/\hbar$. Inset: integrals of the Drude (solid circles) and Lorentz (solid squares) components of the OC, in the low frequency interval of frequencies
$0\le \omega\le {\tilde\omega}/2$. Empty squares are the sum of the two components, that is almost constant with temperature.}
\label{fig_4}
\end{center}
\end{figure}

The ratio of the two coefficients, $\sigma_D / \sigma_L$, quickly increases 
with the temperature in the insulating phase, and saturates in the metallic phase at a value around $8.0$ (Fig.\ \ref{fig_4}). 
Indeed, temperature increasing releases carriers trapped at low temperatures and transfers the spectral weight at low frequencies from the 
Lorentz to the Drude component, which dominates at high temperatures.
In the inset of Fig.~\ref{fig_4} we plot the spectral weight of the two components in the interval $0\le\omega\le {\tilde\omega}/2$, together with
their sum which remains almost equal as changing the temperature, showing that the exchange of the weights between Drude and Lorentz component
is the leading source of the strong temperature dependence of mobility.

Interesting indication of the profound role of disorder even in the high-temperature metallic region follows
from the low-frequency expansion of $\sigma(\omega)$ in powers of $\omega^2$:
\begin{equation}
\sigma(\omega)\simeq \sigma_D + (\sigma_L\tau_L^2-\sigma_D\tau_D^2)\omega^2.
\end{equation}
The sign of the constant before $\omega^2$ is given by the balance between the Drude metallic contribution 
(always negative) and the Lorentz insulating contribution (always positive).
For the high mobility organic materials studied in this work it is found  that the positive contribution 
always overcomes that negative (Fig.\ \ref{fig_4}) at all the studied temperatures.
Positive curvature around $\omega=0$ is very large in the insulating phase due to trapping of carriers,
but it remains positive even at high temperatures, which distinguishes it form typical ordinary metals.

We note that the Lorentz contribution remains nonzero and saturates even at high temperatures where
the majority of trapped carriers are released. 
This persistence of the Lorentz contribution follows from the Feynman polaron scenario, where a carrier 
interacting with a phonon field is described in an effective way as a charge linked, through a spring, 
to a massive particle (simulating the phonon field).
As a  result the system behavior can be decomposed into the free motion of the center of mass of the 
effective quasiparticle (polaron) and the excitations associated to the relative motion of the pair 
(charge - massive particle) which are bound oscillations in the harmonic potential due to the spring.  
In this case one is able to decompose the low frequency behavior of the system into two contributions:
the Drude one, associated with the center of mass motion, and the Lorentz contribution describing the bound oscillations of the internal degrees of freedom
\cite{toyozawa}.
Here we note a substantial change introduced by the disorder into the balance between Drude and Lorentz contributions.
In the crystals without disorder,the Lorentz term is negligible at low temperatures and gains 
a finite weight only at high temperatures  \cite{noi}. 
To the contrary, the Lorentz term dominates at low temperatures in system with notable disorder, because 
bound oscillations around real impurities contribute to the Lorents term much more effectively that those
around the fictitious Feynman charge.

\section{Time dependent diffusivity}
The role of the Drude and Lorentz component in transport can be highlighted studying their contribution to the 
time dependent diffusivity, that is related to the $\sigma(\omega)$ by the following expression \cite{kubo,ciuchi_1,fratini_2}
\begin{equation}
D(t)=\frac{1}{2}\frac{d\Delta x^2}{dt}=\frac{\hbar}{\pi e^2}\int\limits_{0}^{\infty}\frac{\sigma(\omega)\sin(\omega t)}{\tanh(\beta\hbar\omega/2)}d\omega,
\label{Dt}
\end{equation}
where $\beta=1/k_BT$, and $\Delta x^2=\langle[x(t)-x(0)]^2\rangle$ is the mean square displacement of the position operator.

\begin{figure}
\begin{center}
\includegraphics[height=5cm]{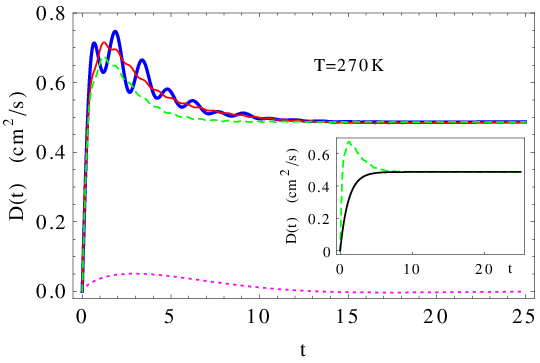}
\caption{Time dependent diffusivity at $T=270K$. Time is measured in terms of $\tau_0=\hbar/t_h$.
Dashed green (dotted magenta) line: contribution of the Drude (Lorentz) term. Red line: sum of the two contributions.
Blue points: full calculation. Inset: comparison between the Drude contribution to the diffusivity in the quantum (dashed green) and classical (solid black) case.
}
\label{fig_5}
\end{center}
\end{figure}

Indeed, the long time behavior is determined solely by the fitted Drude contribution (\ref{eq_Drude}) 
[dashed green curve in Fig.~\ref{fig_5}], whereas the diffusivity due to the fitted Lorentz-term (\ref{eq_Lorentz})
[dotted magenta line] is seen only at small times because localized particles can not contribute into 
long-time large distance diffusion.
Moreover, the Drude term dominates even at small times, showing a cross-over from a quadratic in time 
balistic contribution at very short times to a dissipative behavior at long times going through a maximum 
at a time around $\tau_0$.     
The differences between the full calculations (blue points) and the sum of the fitted Drude and 
Lorentz contributions (solid red curve) are due to the high frequencies oscillations obtained in the 
full calculations, that are not reproduced by the sum of Drude and Lorentz contributions. 
We note that the diffusivity at long times (mobility) is characterized by a scattering time 
$\tau_D\simeq \tau_0\simeq10^{-15}\,\mbox{s}$, which is few times smaller than 
that typical of good metals. 
In this regime the Drude term gives rise to a peak in the time depending diffusivity at a time around $\tau_0$
(see dashed green curve in  Fig.~\ref{fig_5}) signaling the quantum nature of the transport.
Indeed, in the classical case, given by Eq.~(\ref{Dt}) in the limit $\hbar\to 0$,
the time dependent diffusivity would be a monotonically increasing function of time, as shown by the solid black curve in Fig.~\ref{fig_5}.

It is worth to note that correlations and memory effects have been extensively studied in other contexts, in particular in statistical physics and studies
of dieletric polarization relaxation \cite{dielectric}.

\section{Conclusions}
We suggested a model, where transport properties originate from the interplay of two different interactions.
Charge carriers are coupled to phonons by SSH-type interaction, and scattered by quenched local impurities. 
Within the model proposed by Troisi and Orlandi \cite{troisi_1}, our approximation free results
provide quantitative theoretical description of the conductivity of  high 
mobility organic crystals Pentacene and Rubrene from the insulating low- to the metallic high-temperature
regime.
Essential ingredient of our model is the coupling to impurities which was missing in previous approaches. 
We are not aware of any other results giving theoretical description of transport data for 
such wide temperature range. 

The scenario emerging from our calculations is based on the existence of two additive contributions
to the optical conductivity, characterized by two different times. One ($\tau_D$) controls the Drude contribution, and is of the order of $\hbar/t_h$.
It is the time when the diffusive motion sets in, signaling the onset of a fully incoherent motion, whereas for shorter times the motion is of course ballistic.
This contribution is attributed to the motion of the center of mass of the relevant quasi-particle.
The other contribution is characterized by a longer time scale ($\tau_L$), and controls the maximum of the optical conductivity ($\omega_L\simeq 1/\tau_L$).
This time is related to the relative motion of the internal degrees of freedom of the quasi-particle.

We revealed several fingerprints which discern transport in the high mobility organic crystals investigated in this paper, from that
in ordinary metals: extremely small scattering times, diffusion length comparable to lattice spacing, 
and substantial contribution of localized charges to the high temperature optical conductivity. 

\acknowledgments

This work was supported by the ImPACT Program of the Council for Science, Technology and
 Innovation (Cabinet Office, Government of Japan).

\bibliography{paper}{}
\bibliographystyle{apsrev4-1}

\onecolumngrid
{}

\clearpage
\begin{center}
\bf\large Supplemental material 
\end{center}

\setcounter{section}{0}
\twocolumngrid
\section{Optical conductivity}

Within the linear response theory,  the optical conductivity (OC) \cite{sha}, $\sigma(z)$, can be obtained calculating the current-current correlation function:
\begin{eqnarray}
\sigma(z)=\frac{i}{z}\left(\Pi(z)- \Gamma \right),
\label{sigma}
\end{eqnarray}
where $z$ lies in the complex upper half-plane, $z=\omega+i0^+$. The quantity $\Gamma$ is:
\begin{eqnarray}
\Gamma= -\frac{1}{\hbar}\int_0^{\beta\hbar} ds  \left\langle j(s)j(0) \right\rangle,
\label{gam}
\end{eqnarray}
and $\Pi(z)$ represents the current-current
correlation function
\begin{eqnarray}
\Pi(z)=-\frac{i}{\hbar} \int_0^{\infty} d\tau e^{i z \tau} \left\langle [j(\tau),j(0)] \right\rangle.
\label{pi}
\end{eqnarray}

In Eq.~(\ref{pi}) (Eq.~(\ref{gam})) $j(\tau)$ ($j(s)$) is the real-time (imaginary-time)
Heisenberg representation of the
current operator, $[$,$]$ denotes the commutator, 
$\left\langle \right\rangle$ indicates the thermal average, and $\beta=1/k_BT$.
The real part of OC 
is related to the 
imaginary-time current-current correlation function: 
\begin{eqnarray}
\Pi(s)=\int_{-\infty}^{\infty} d\omega \frac {1}{\pi} \frac {\omega e^{-\omega s}}
{1-e^{\beta\hbar\omega}} \operatorname{Re} \sigma(\omega). 
\label{analytic}
\end{eqnarray}

The function $\Pi(s)$ has been calculated by using World Line Monte Carlo \cite{antonio,noi}
methods. The dynamical spectra, then, 
is extracted  from the integral equation, Eq.~(\ref{analytic}), through two different approaches i) the maximum entropy method \cite{JM}
and ii) the stochastic optimization method \cite{andrey}.
In the former case, in particolar, we followed the Bryan's method choosing, as default model, the OC obtained through exact diagonalization
on a lattice of $20$ sites with periodic boundary conditions  (at the investigated temperatures the mean free path (MFP) is
less than $6a$, so that such a small lattice provides a very good starting point). 

\section{Sum rule for optical conductivity}

\begin{figure}[ht]
\begin{center}
\includegraphics[width=5cm]{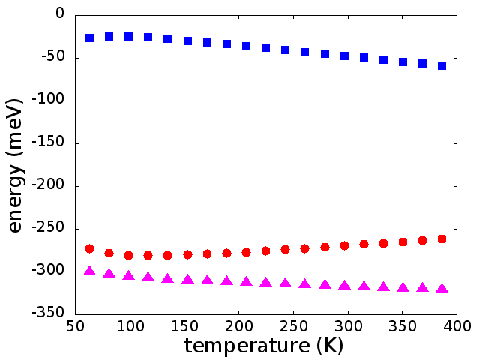}
\caption{Thermal average of the kinetic energy (red circles), electron phonon interaction (blue squares), and their sum (magenta triangles).}
\label{fig_1}
\end{center}
\end{figure}

The optical conductivity obeys the sum rule
\begin{equation}
\int\limits_0^\infty\sigma(\omega)d\omega=
-\frac{\pi}{2}\left(\frac{ea}{\hbar}\right)^2 \langle H_2 \rangle \; ,
\label{suru} 
\end{equation}
where $\langle H_2 \rangle$ is the thermal average of $H_2=H_K+H_{EPI}$, and 
\begin{align}
H_K &= \sum_{q}\varepsilon(q)c_q^\dag c_q,
\\
H_{EPI}&=\sum_{q,k} M(k,q) \left(a_{-k}^\dag+a_{k}\right)c_{q+k}^\dag c_q \;,
\end{align}
are the  kinetic energy of the electron, and the electron-phonon interaction
(see main text for the definition of $\varepsilon(q)$ and $M(k,q)$).
In Fig.~\ref{fig_1} we show the thermal average of $H_K$ (red circles), $H_{EPI}$ (blue squares),
and the sum $H_2$ (magenta triangles).
The average $\langle H_2 \rangle$ is almost constant in the considered temperature range, while there is an exchange
between the two components $\langle H_K \rangle$ and $\langle H_{EPI} \rangle$.

\section{Adiabatic approach}

\begin{figure}[ht]
\begin{center}
\includegraphics[width=5cm]{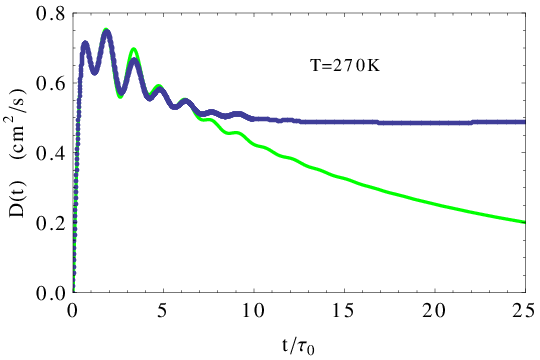}
\caption{The time depending diffusivity in the approximation free MonteCarlo calculations (blue line), and in the adiabatic approximation (green line) at $T=270\,\mbox{K}$. }
\label{fig_2}
\end{center}
\end{figure}

In Fig.~\ref{fig_2} we compare time dependent diffusivity given by the Monte Carlo full calculation (blue line),
with the so-called adiabatic limit (green line), at temperature $T=270\,\mbox{K}$. In the adiabatic approach the quantum phonon dynamics is completely neglected 
and they are approximated as static lattice deformations \cite{cataudella_1}. As it can be noted the adiabatic result closely follows the full result at short 
times ($t<7\,\tau_0$, where $\tau_0$ is $\hbar$ divided by the hopping parameter) but then it deviates tending to zero as it is expected by the Anderson theorem. 

We emphasize that, within our approach, the numerical exact solution is characterized by two times, $\tau_D$ controlling the Drude contribution, which is always close to $\tau_0$, and 
$\tau_L$, controlling the localized Lorentz contribution which assumes larger values ranging from $16.2\,\tau_0$ at $T=124\,\mbox{K}$ to $4\,\tau_0$ at $T=270\,\mbox{K}$.
It is worth noting that the time emerging from Fig.~\ref{fig_2}, the time when the full calculation deviates from the fully adiabatic results, does not coincide neither with $\tau_D$, nor
with $\tau_L$.


\end{document}